\documentclass[12pt]{iopart}
\begin{document}
\title{Study of a confined Hydrogen-like atom by the Asymptotic Iteration Method}
\author{Hakan Ciftci}
\address{
Gazi \"{U}niversitesi, Fen-Edebiyat Fak\"{u}ltesi, \\
Fizik B\"{o}l\"{u}m\"{u}, 06500, Teknikokullar, Ankara, Turkey
}\ead{hciftci@gazi.edu.tr}
\author{Richard L. Hall}
\address{Department of Mathematics and Statistics, Concordia University,
1455 de Maisonneuve Boulevard West, Montr\'eal,
Qu\'ebec, Canada H3G 1M8} \ead{rhall@mathstat.concordia.ca}
\author{Nasser Saad}
\address{Department of Mathematics and Statistics,
University of Prince Edward Island,
550 University Avenue, Charlottetown,
PEI, Canada C1A 4P3.}\ead{nsaad@upei.ca}
\def\dbox#1{\hbox{\vrule  
                        \vbox{\hrule \vskip #1
                             \hbox{\hskip #1
                                 \vbox{\hsize=#1}%
                              \hskip #1}%
                         \vskip #1 \hrule}%
                      \vrule}}
\def\qed{\hfill \dbox{0.05true in}}  
\def\square{\dbox{0.02true in}} 
\begin{abstract}
The asymptotic iteration method (AIM) is used to obtain both special exact solutions and general approximate solutions for a Hydrogen-like atom confined in a spherical box of arbitrary radius $R$. Critical box
radii, at which states are no longer bound, are also calculated. The results are compared with those in the literature.
\end{abstract}

\vspace{2pc}

\maketitle
\section{Introduction}
Recently, there has been great interest in studying the properties of confined quantum systems. There are many reasons for this. The recent developments in nanotechnology have generated intensive research activity in modeling spatially confined quantum systems \cite{Michels}-\cite{colm}.  When an atom or a molecule is trapped inside any
kind of microscopic cavity, or is placed in a high pressure environment, it experiences spatial confinement
that affects its physical and chemical properties \cite{Sen}-\cite{Sinha}. The concept of a confined quantum system goes back to the early work of Michels et al \cite{Michels} who studied the properties of an atomic system under very high pressures. They suggested to replace the interaction of the
atoms with surrounding atoms by a uniform pressure on a sphere within
which the atom is considered to be en closed. This led them to consider
the problem of hydrogen with modified external boundary conditions \cite{colm}. Since then, the confined hydrogen atom attracted widespread attention \cite{colm}-\cite{Cecil}.  The approach to this confined problem is the same as that used to study a hydrogenic donor located at the center of a spherical $GaAs - (Ga,Al)As$
quantum dot, a semiconductor device that confines electrons \cite{Zhu}-\cite{Varshni}.  
Considerable interest in calculating the properties of donor states in a quantum dot has been renewed recently and a number of calculations on the bound states of a hydrogenic donor in a quantum dot have been reported (\cite{Varshni}, and the references therein). Many researchers have
carried out accurate calculations of eigenvalues of the confined hydrogen atom using various techniques. Some of these
are variational methods \cite{Gimarc}-\cite{ypv}, finite element methods \cite{Guimaraes}, and
algebraic methods \cite{Burrows}. In the present work, we have calculated the
energy eigenstates of the confined hydrogen atom for different quantum
levels using the Asymptotic Iteration Method (AIM) \cite{Ciftci}. In addition to calculating the energy eigenvalues, we also used AIM to compute the critical cage radius at which a state is no longer bound.
Sommerfeld and Welker \cite{Sommerfeld} gave a detailed investigation on the variation of
the binding energy of $1s$ state of hydrogen atom with respect to the sphere
radius. They showed that there is critical value of the sphere radius $r_c$ at
which the binding energy is zero. This radius is now known as the `critical cage
radius' \cite{Var}. It has been found that,
for $r<r_{c}$, the energy of the system is positive, i.e.  the electron exerts pressure on the walls of the sphere \cite{Var}. Later some workers
 calculated this critical cage radius for various atoms
and different quantum levels. For example, Boeyens \cite{Boeyens} has calculated the $%
r_{c}$ values for the ground state of many atoms by the
Hartree-Fock-Slater method.

The organization of the paper is as follows. In the next section, we give a brief outline of the confined Hydrogen-like atom problem. In section 3, we summarize the asymptotic
iteration method (AIM) used in this work. In section 4, we apply AIM to the confined Hydrogen-like atom problem and derive certain exact solutions for special parameter values. In section~5, we present the results of our general calculations for any box radius $R$, and we compare them with the most accurate results available in the literature. We present our calculated critical cage radii in section~6.
\section{Formulation of the Problem}
In atomic units, the radial Schr\"{o}dinger equation of a Hydrogen-like atom located in the
center of a spherical box of radius $R$ can be written as
\begin{equation}\label{eq21}
\left[ -\frac{d^{2}}{dr^{2}}+\frac{l(l+1)}{r^{2}}-\frac{A}{r}\right] \psi
(r)=E\psi (r),~\psi (0)=\psi (R)=0
\end{equation}
where $A>0$ is real parameter, $E$ is the atom's energy and the boundary condition for the exact eigenfunction $\Psi(r)$ is 
$\Psi(0)=\Psi(R)=0$ (i.e. it satisfies the Dirichlet boundary condition). Here, ${\mathbf r}$ is the electron position vector with respect to the nucleus and $r=||{\mathbf r}||$ is the length of this vector. Of course, the presence of the Dirichlet boundary condition changes the structure of the atomic energy spectrum drastically as compared with that  of the free, non-confined, hydrogen atom  corresponding to the limit $R\rightarrow \infty$.  The interesting hidden symmetry that manifests itself in extra degeneracy of energy levels at specific values of $R$ has been studied \cite{sch}-\cite{pupy} in number of articles. 

\noindent It is interesting to note that when $r$ tends to zero, the wavefunction $\Psi(r)$ behaves as $r^{l+1}$, and when 
$r$ approaches $R$, $\Psi $ must approach zero. Under these assumptions, we may represent the unnormalized wavefunction in the form
\begin{equation}\label{eq22}
\psi(r) = \left\{ \begin{array}{ll}
 r^{l+1}(R-r)\exp (-ar)f(r), &\mbox{if $r<R$} \\
  0 &\mbox{if $r\geq R$}
       \end{array} \right.
\end{equation}
where $f(r)$ is to be determined, and $a$ is a parameter that will be used to obtain the energy and to control the convergence of the iterative method used by AIM. As we shall show, this simple form of wave function (\ref{eq22}) gives good results for the confined hydrogen atom. Substituting (\ref{eq22}) in the Schr\"odinger equation (\ref{eq21}), we find that the radial function $f(r)$ must satisfy the differential equation
\begin{eqnarray}\label{eq23}
f^{\prime \prime }(r)&=&2\left( a+\frac{1}{R-r}-\frac{l+1}{r}\right) f^{\prime
}(r)\nonumber\\
&+&\left( \frac{(2l+2)a-A}{r}+\frac{2l+2}{r(R-r)}-\frac{2a}{R-r}\right)
f(r)
\end{eqnarray}
where we denote 
\begin{equation}\label{eq24}
E=-a^{2}.
\end{equation}
The problem is then reduced to solving this second-order homogeneous differential equation for $f(r).$ The asymptotic iteration method (AIM) was developed with the idea of using minimal algebraic
computations to solve such differential equations. In the next section, we
give a brief introduction to the asymptotic iteration method. Detailed proofs and applications to unconfined systems can
be found in \cite{Ciftci} and \cite{Hakan}.
\section{Brief Introduction to the Asymptotic Iteration Method}
\medskip Given $\lambda _{0}(x)$ and $s_{0}(x)$ sufficiently differentiable functions, the asymptotic iteration method tells us that 
the second-order
differential equation 
\begin{equation}\label{eq31}
y^{\prime \prime }=\lambda _{0}(x)y^{\prime }+s_{0}(x)
\end{equation}
has a general solution
\begin{equation}\label{eq32}
y(x)=\exp(-\int^{x}\alpha dt)\left( C_{2}+C_{1}\int^{x}\exp \left(
\int^{t}(\lambda _{0}(\tau )+2\alpha (\tau ))d\tau \right) dt\right)
\end{equation}
if for some $n>0$,
\begin{equation}\label{eq33}
{\frac{s_{n}}{\lambda _{n}}}={\frac{s_{n-1}}{\lambda _{n-1}}}\equiv \alpha
\end{equation}
where
\begin{equation}\label{eq34} \left\{ \begin{array}{ll}
\lambda _{n}=\lambda _{n-1}^{\prime }+s_{n-1}+\lambda _{0}\lambda
_{n-1}, &\mbox{} \\
  s_{n}=s_{n-1}^{\prime }+s_{0}\lambda _{n-1}.
 &\mbox{}
       \end{array} \right.
\end{equation}
\vskip0.1true in
In general, the (asymptotic) termination condition Eq.(7) can be
written equivalently as follows
\begin{equation}\label{eq35}
{\delta_n (r)}=\lambda _{n}s_{n-1}-s_{n}\lambda _{n-1}=0,\quad n=1,2,\dots
\end{equation}
Note that, we can start the computation of the recurrence relation (\ref{eq35}) from $n=0$ with the initial conditions $\lambda_{-1}=1$ and $s_{-1}=0$. It follows that if $\delta_n(r)=0$, then $\delta_{n+1}(r)=0$ for all $n$. The termination condition (\ref{eq35}) has a crucial role for the computation of the eigenenergies (\ref{eq24}). Indeed, using (\ref{eq23}), we obtain, by means of
\begin{equation}\label{eq36} \left\{ \begin{array}{ll}
\lambda_0=2( a+\frac{1}{R-r}-\frac{l+1}{r}), &\mbox{} \\
s_0=( \frac{(2l+2)a-A}{r}+\frac{2l+2}{r(R-r)}-\frac{2a}{R-r}),
 &\mbox{}
       \end{array} \right.
\end{equation}
the recursive relations for  $s_{n}$ and $\lambda _{n}$, $n=1,2,\dots$ given by (\ref{eq34}). These quantities are, in general, functions of the parameter $a$ and the variable $r$. If, for
a suitable choice of $a$, the termination condition Eq.(\ref{eq35}) is satisfied at every $r<R$,
then the problem is called `exactly solvable'. If the differential equation is not exactly
solvable, the function $\delta (r)$, in general, will depends on both $a$ and $r$. In this
case, we obtain $a$ by iterating Eq.(\ref{eq35}) with a suitable initial value of $r=r_{0}<R$.
\section{Exact analytical solutions}
In this section, we first present the exact analytical solutions 
for a Hydrogen-like atom confined in a spherical box of radius $R$. 
The application of AIM which we employ here shows that for certain values of
the parameter $a$, we obtain exact solution provided the confinement radius $R$ assumes 
corresponding definite values; that is to say, Eq.(\ref{eq23}) is then exactly solvable.
We first note that the integer $n$ here is not to be identified for the hydrogenic problem with
the radial quantum number, but rather by $n = 1 + $number of radial nodes.  Thus, with this notation,  in the limit as $R\rightarrow \infty,$ the eigenvalues of Eq.(1) become exactly $E = -A^2/(4(n+\ell)^2).$ Using (\ref{eq35}) and (\ref{eq36}) we find the following results.
\begin{itemize}
\item For $n=1$ and $a={A\over 2(l+2)}$, we have
 $\delta_1=0$ if
\begin{equation*}
AR-2(l+1)(l+2)=0.
\end{equation*}
\item For $n=2$ and $a={A\over 2(l+3)}$, we have $\delta_2=0$ if
\begin{equation*}
A^2R^2-2(2l+3)(l+3)AR +2(l+3)^2(2l+3)(l+1)=0.
\end{equation*}
\item For $n=3$ and $a={A\over 2(l+4)}$, we have $\delta_3=0$ if
\begin{eqnarray*}
A^3R^3&-&6(l+2)(l+4)A^2R^2 +6(l+4)^2(l+2)(2l+3)AR\\
&-&4(l+4)^3(l+2)(l+1)(2l+3)=0.
\end{eqnarray*}
\end{itemize}
and so on. In general, we have for
\begin{equation}\label{eq41}
{a=}\frac{A}{2(n+l+1)},\quad l=0,1,2,\dots
\end{equation}
that $\delta_n=0$,~ $n=1,2,\dots$ if
\begin{equation}\label{eq42}
\sum\limits_{k=0}^{n}\frac{(-1) ^{k}(l+n+1)^{n-k}\Gamma(2l+n+2)\Gamma(n+1) }{%
\Gamma(2l+k+2)\Gamma(n+1-k)\Gamma(k+1)}
(AR)^{k}=0.
\end{equation}
It is interesting to note that the polynomial conditions on $R$ given by (\ref{eq42}) can be written in terms of the confluent hypergeometric functions as
\begin{equation}\label{eq43}
{}_1F_1\left(-n; 2l+2;\frac{AR}{l+n+1}\right)=0, \quad \quad n=1,2,\dots.
\end{equation}
This means that, in order to obtain the energy spectrum as given by (\ref{eq24}), we must find the roots of the confluent hypergeometric function (\ref{eq43}).
In this case, for given $A$,  the eigenvalues (\ref{eq24}) are given by  
\begin{equation}\label{eq44}
E_{nl}=-\frac{A^{2}}{4(n+l+1)^{2}}, \quad n=1,2,\dots,~l=0,1,2,\dots
\end{equation}
where $R$ has values which are the roots of (\ref{eq43}). The corresponding analytic wave functions can be computed by using Eq.(\ref{eq32}) for $A$ and $R$, related by means of (\ref{eq43}) for given $n$ and $l$.
Straightforward computations then show that
\begin{equation}\label{eq45}
f_{1l}^{0}(r)=1,\quad f_{nl}^{m}(r)=\prod\limits_{i=0}^{n-1}\left( 1-\frac{r}{_{R_{i}}}\right),~i\neq m
\end{equation}
where $m=0,1,2,...n-1$ and $R_i$ are the roots of (\ref{eq43}). In Table 1, we report some eigenenergies of the enclosed hydrogen atom as a function of the radius $R$ along with the corresponding wave functions $f(r)$ computed by using (\ref{eq45}).
\begin{table}[ht]
\caption{Exact energy eigenvalues for certain fixed values of $R$.}\medskip 
\centering 
\begin{tabular}{|l|l|l|l|l|l|}
\hline
\hline
$n$ & $l$ & $m$ & $R$ & $E$ & $f_{nl}^{m}(r)$ \\ \hline
$1$ & $0$ & $0$ & $\frac{4}{A}$ & $-\frac{A^{2}}{16}$ & $1$ \\ \hline
$1$ & $1$ & $0$ & $\frac{12}{A}$ & $-\frac{A^{2}}{36}$ & $1$ \\ \hline
$1$ & $2$ & $0$ & $\frac{24}{A}$ & $-\frac{A^{2}}{64}$ & $1$ \\ \hline
$1$ & $3$ & $0$ & $\frac{40}{A}$ & $-\frac{A^{2}}{100}$ & $1$ \\ \hline
$2$ & $0$ & $0$ & $\frac{3(3-\sqrt{3})}{A}$ & $-\frac{A^{2}}{36}$ & $1-\frac{%
Ar}{3(3+\sqrt{3})}$ \\ 
~ & ~ & $1$ & $\frac{3(3+\sqrt{3})}{A}$ & $-\frac{A^{2}}{36}$ & $1-\frac{%
Ar}{3(3-\sqrt{3})}$ \\ \hline
$2$ & $1$ & $0$ & $\frac{4(5-\sqrt{5})}{A}$ & $-\frac{A^{2}}{64}$ & $1-\frac{%
Ar}{4(5+\sqrt{5})}$ \\ 
~ & ~ & $1$ & $\frac{4(5+\sqrt{5})}{A}$ & $-\frac{A^{2}}{64}$ & $1-\frac{%
Ar}{4(5-\sqrt{5})}$ \\ \hline
$2$ & $2$ & $0$ & $\frac{5(7-\sqrt{7})}{A}$ & $-\frac{A^{2}}{100}$ & $1-%
\frac{Ar}{5(7+\sqrt{7})}$ \\
~ & ~ & $1$ & $\frac{5(7+\sqrt{7})}{A}$ & $-\frac{A^{2}}{100}$ & $1-%
\frac{Ar}{5(7-\sqrt{7})}$ \\ \hline
$2$ & $3$ & $0$ & $\frac{36}{A}$ & $-\frac{A^{2}}{144}$ & $1-\frac{Ar}{72}$
\\ 
~ & ~ & $1$ & $\frac{72}{A}$ & $-\frac{A^{2}}{144}$ & $1-\frac{Ar}{36}$
\\ \hline
$3$ & $0$ & $0$ & $\frac{3.74329}{A}$ & $-\frac{A^{2}}{64}$ & $( 1-%
\frac{Ar}{13.2216})( 1-\frac{Ar}{31.0351}) $ \\
~ & ~ & $1$ & $\frac{13.2216}{A}$ & $-\frac{A^{2}}{64}$ & $( 1-%
\frac{Ar}{3.74329}) ( 1-\frac{Ar}{31.0351}) $ \\ 
~ & ~ & $2$ & $\frac{31.0351}{A}$ & $-\frac{A^{2}}{64}$ & $( 1-%
\frac{Ar}{13.2216}) ( 1-\frac{Ar}{3.74329}) $ \\ \hline
\end{tabular}
\end{table}
\bigskip
We can obtain the complete wavefunction from Eq.(\ref{eq22}).  For given $n$, $l$, $m$
quantum numbers, the corresponding (un-normalized) wave functions are given by
\begin{equation}\label{eq46}
\Psi _{nl}^{m}=N_{nl}^m ~r^{l+1}\exp \left( -\frac{A}{2\left( n+l+1\right) }r\right)
\prod\limits_{i=0}^{n-1}\left( 1-\frac{r}{_{R_{i}}}\right),~r\in
\left( 0,R_{m}\right) 
\end{equation}
where $R_{i(m)}$ are given as the roots of the confluent hypergeometric function Eq.(\ref{eq43}). We have first obtained these exact solutions of the confined hydrogen-like problem, valid for special values of the parameters, since they can be very useful for verifying the correctness of general approximations.  In the next section, we will give more general results for the problem by applying AIM to Eq.(\ref{eq23}) for arbitrary given $R$.
\section{Energy eigenvalues of a Hydrogenic Donor}
The Hamiltonian of an on-center impurity in a spherical quantum dot can be written in
the effective-mass approximation as
\begin{equation}\label{eq51}
H = -{\hbar^2\over 2m^*}\nabla^2-{e^2\over \epsilon r}
\end{equation}
where $m^*$ is the effective mass and $\epsilon$ is the dielectric constant of the material of the quantum
dot. The donor is assumed to be at the center of the quantum dot of radius $R$ with an infinite
barrier height. This means that the wave function vanishes at $r = R$.
In atomic units, the radial Hamiltonian equation for the Coulomb Potential is given by 
\begin{equation}\label{eq52}
\left[ -\frac{d^{2}}{dr^{2}}+\frac{l(l+1)}{r^{2}}-\frac{2}{r}\right] \psi
(r)=E\psi (r),\quad \psi (0)=\psi (R)=0.
\end{equation}
In this section, we use AIM in order to obtain the energy states of the confined atom. We note first that, Eq.(\ref{eq23}) now reads
\begin{eqnarray}\label{eq53}
f^{\prime \prime }(r)&=&2\left( a+\frac{1}{R-r}-\frac{l+1}{r}\right) f^{\prime
}(r)\nonumber\\
&+&\left( \frac{(2l+2)a-2}{r}+\frac{2l+2}{r(R-r)}-\frac{2a}{R-r}\right)
f(r)
\end{eqnarray}
Now, for a given $R$ value, we have calculated the energy values $a$ using (\ref{eq35}) and the recursive relations (\ref{eq34}) initiated with 
\begin{equation}\label{eq54} \left\{ \begin{array}{ll}
\lambda_0=2( a+\frac{1}{R-r}-\frac{l+1}{r}), &\mbox{} \\
\\
s_0=( \frac{(2l+2)a-2}{r}+\frac{2l+2}{r(R-r)}-\frac{2a}{R-r}).
 &\mbox{}
       \end{array} \right.
\end{equation}
Since the system is not exactly solvable, and since $\delta_n(r)\equiv\delta_n(a;r)=0$, we must
choose a suitable value for $r_{0}<R$ in order to initiate AIM. For our numerical results, we have fixed $r=r_{0}$ as 
$\frac{R}{2}$ for $R\leq 1$ and $1$ for $R\geq 1$. The corresponding results are shown in Tables 2 and 3.  Comparison is made with results from \cite{Varshni}, which are based on a variational method.
\begin{table}[ht]
\caption{Exact energy eigenvalues for the 1st state (in Rydbergs) found with AIM, for different values of $R$. Comparison is made with the results of Ref. \cite{Varshni}.}
\medskip 
\centering 
\begin{tabular}{|l|l|l|l|}
\hline
\hline
$R$ & $a$ & $E_{AIM}$ & $E_{exact}$\cite{Varshni} \\ \hline
$0.1$ & $30.626~558~365~553~640~428~52i$ & $937.986~077~318~663~675~020$ & $937.986$ \\ \hline
$0.2$ & $14.904~352~306~411~647~880~03i$ & $222.139~717~673~638~207~696$ & $222.14(0)$ \\ \hline
$0.3$ & $~9.653~226~131~767~727~449~50i$ & $~93.184~774~751~043~322~516$ & $93.185(0)$ \\ \hline
$0.4$ & $~7.019~089~585~470~665~068~48i$ & $~49.267~618~608~862~752~786$ & $49.268(0)$ \\ \hline
$0.5$ & $~5.431~016~485~033~032~864~57i$ & $~29.495~940~060~700~559~289$ & $29.496(0)$ \\ \hline
$0.6$ & $~4.365~250~922~031~022~986~22i$ & $~19.055~415~612~292~696~322$ & $19.055(0)$ \\ \hline
$0.7$ & $~3.597~200~613~602~544~685~31i$ & $~12.939~852~254~502~523~992$ & $12.940(0)$ \\ \hline
$0.8$ & $~3.014~425~378~412~749~608~39i$ & $~~9.086~760~362~018~848~673$ & $9.0868(0)$ \\ \hline
$0.9$ & $~2.554~286~411~599~184~224~80i$ & $~~6.524~379~072~480~237~168$ & $6.5244(0)$ \\ \hline
$1.0$ & $~2.178~986~400~188~704~127~38i$ & $~~4.747~981~732~207~327~454$ & $4.7480(0)$ \\ \hline
$1.2$ & $~1.593~307~889~331~964~142~70i$ & $~~2.538~630~030~207~478~496$ & $2.5386(0)$ \\ \hline
$1.4$ & $~1.137~633~609~176~779~056~96i$ & $~~1.294~210~228~728~584~474$ & $1.2942(0)$ \\ \hline
$1.6$ & $~0.736~630~589~422~075~466~49i$ & $~~0.542~624~625~272~314~320$ & $0.54262$ \\ \hline
$1.8$ & $~0.255~171~521~889~990~724~15i$ & $~~0.065~112~505~583~654~015$ & $0.06511$ \\ \hline
$2.0$ & $~0.5$ & $-0.25^*$ & $-0.25$ \\ \hline
$2.2$ & $~0.681~224~443~905~427~229~15$ & $-0.464~066~742~974~258~570$ & $-0.46407$ \\ \hline
$2.4$ & $~0.782~812~885~873~172~513~99$ & $-0.612~796~014~289~084~615$ & $-0.61280$ \\ \hline
$2.6$ & $~0.847~323~189~868~271~355~52$ & $-0.717~956~588~088~542~630$ & $-0.71796$ \\ \hline
$2.8$ & $~0.890~692~535~878~441~133~67$ & $-0.793~333~193~469~568~146$ & $-0.79333$ \\ \hline
$3.0$ & $~0.920~833~630~721~047~974~35$ & $-0.847~934~575~466~907~348$ & $-0.84793$ \\ \hline
$3.2$ & $~0.942~234~555~027~191~412~81$ & $-0.887~805~956~687~289~402$ & $-0.88781$ \\ \hline
$3.4$ & $~0.957~650~893~749~316~017~63$ & $-0.917~095~234~298~863~756$ & $-0.91710$ \\ \hline
$3.6$ & $~0.968~866~519~413~519~397~41$ & $-0.938~702~332~440~467~559$ & $-0.93870$ \\ \hline
$3.8$ & $~0.977~080~758~210~929~418~85$ & $-0.954~686~808~066~044~717$ & $-0.95469$ \\ \hline
$4.0$ & $~0.983~122~883~548~157~499~83$ & $-0.966~530~604~156~044~052$ & $-0.96653$ \\ \hline
\end{tabular}\\
$^{*}$ Exact, see Table 1
\end{table}

\begin{table}[ht]
\caption{Exact eigenenergies (in Rydbergs) obtained with AIM for different values of $R$ for the $2p$ state ($\ell=1$). Comparison is made with results of Ref. \cite{Varshni}, see also Ref. \cite{ypv}.}\medskip 
\centering 
\begin{tabular}{|l|l|l|l|}
\hline
\hline
$R$ & $a$ & $E_{AIM}$ & $E_{exact}$~\cite{Varshni} \\ \hline
$0.4$ & $10.811~856~798~907~243~508~22i$ & $116.896~247~440~076~786~588~403$ & $116.896$ \\ \hline
$1$ & $~4.055~400~921~280~376~683~83i$ & $~~16.446~276~632~321~727~964~741$ & $~16.446$ \\ \hline
$2$ & $~1.775~397~862~793~768~780~65i$ & $~~~3.152~037~571~212~681~836~807$ & $~3.1520$ \\ \hline
$4$ & $~0.535~774~362~421~267~801~33i$ & $~~~0.287~054~167~427~916~019~155$ & $~0.28705$ \\ \hline
$8$ & $~0.457~055~940~572~194~782~44$ & $-0.208~900~132~812~333~648~630$ & $-0.2089$ \\ \hline
\end{tabular}
\end{table}

\section{Critical cage radii}
In this section we report the results of our calculations for the critical cage
radius, at which the total energy is zero, i.e., when
the kinetic and potential energy contributions cancel
one another. Such cage radii were first noted by Sommerfeld and Welker \cite{Sommerfeld} in their detailed study on the variation of the binding energy of the $1s$ state of the hydrogen atom, as a function of the sphere radius,
$R$. Assuming that the surface of the spherical box is impenetrable, they showed
that, as $R$ decreases, the binding energy diminishes, and there is a critical value of the
sphere radius at which the binding energy becomes zero. More systematic studies were later carried by Varshni \cite{Var} where the critical cage radius was first recorded; see also \cite{Varsh}. 
In this section we shall use AIM to calculate the cage radius at which the eigenvalues
are zero. With $E=0$, we find, by using (\ref{eq24}), that $a=0$; and then Eq.(\ref{eq53}) reduces to 
\begin{equation}\label{eq61}
f^{\prime \prime }(r)=2\left( \frac{1}{R-r}-\frac{\ell+1}{r}\right) f^{\prime
}(r)+\left( -\frac{2}{r}+\frac{2\ell+2}{r(R-r)}\right) f(r)
\end{equation}

\noindent We now apply AIM with $\lambda_0$ and $s_0$ extracted from (\ref{eq61}). The termination condition $\delta_n(r)=0$,  is then dependent (see (\ref{eq35})) on both the variable $r$ and the parameter $R\equiv r_c$. Starting with $r\approx R/2$, the iteration process converges quickly to the cage radius $R\approx r_{c}$. For example, if $\ell=0$ and $n=5$, AIM takes 26 iterations to yield $r_c=48.09774$ initiated with starting value of $r=24$. The calculated
values of $r_c$ found by using AIM are shown in Table 4 for different values of $l$ and $n$. Similar tables can be easily constructed for arbitrary values of $n$ and $\ell$. The numerical computation of eigenvalues and the critical cage radii in tables 2-4 were performed by using Maple version 10 running on an IBM architecture personal computer (Dell Dimension 4400). In many cases, we have removed some of the apparent divergence \cite{champ} experienced by AIM by increasing the number of significant digits that Maple uses in numerical computation; for the present results we have used \emph{Digits}=50. In order to accelerate the computation we have written our code for the root-finding algorithm, instead of using the default procedure of Maple. All the numerical results reported using AIM in tables 2-4 are exact in sense that the numerical integration of the corresponding Schr\"ondinger (confined) equation yields the same values.

\begin{table}[ht]
\caption{Exact values  of the critical cage radius calculated by the use of AIM $r_{c}(AIM)$. Radii
are in units of the Bohr radius. Comparison with \cite{Var} are also reported $r_{c}$.}\medskip
\centering 
\begin{tabular}{|l|l|l|l||l|l|l|l|}
\hline
\hline
$\ell$ & $n$ & $r_{c}(AIM)$ & $r_{c}$ &$\ell$ & $n$ & $r_{c}(AIM)$ & $r_{c}$ \\ \hline
$0$ & $1$ & $~1.835~246~330~265~5$ & $1.8352$ & $1$ & $1$ & $~5.088~308~227~275~0$ & $5.0883$ \\ \hline
& $2$ & $~6.152~307~040~211~8$ & $6.1523$ & $~$ & $2$ & $11.909~696~568~004~6$ &$11.910$\\ \hline
& $3$ & $12.937~431~736~892~1$ & $12.937$ & $~$& $3$ & $21.174~431~228~262~4$ & $21.174$ \\ \hline
& $4$ & $22.190~095~851~725~6$ & $22.190$ & $~$& $4$ & $32.900~106~781~876~0$ & $32.900$ \\ \hline
& $5$ & $33.910~206~784~109~2$ & $33.910$ & $~$& $5$ & $47.090~674~929~020~9$ & $47.091$ \\ \hline
& $6$ & $48.097~738~137~838~7$ & $48.098$ & $~$&$6$ & $63.747~459~484~409~4$ & $63.747$ \\ \hline
\end{tabular}
\end{table}

\section{Conclusion}
In this work, we have applied AIM to obtain the energy eigenstates of a confined hydrogen atom. The problem is similar to the confinement of electrons in a quantum dot. For certain cases, the  eigenvalues under confinement are given by the roots of the Kummer (confluent) hypergeometric functions whose analytic and numerical properties are well
known \cite{Abramowitz}. The method can be easily adapted to the
study of more highly excited states, without necessitating extensive further algebraic manipulation or numerical work \cite{Cecil}.
We have also calculated the critical cage radii $r_c$ at which various states become unbound. The numerical results are compared with the most accurate results in the literature. It is worth pointing that AIM provides a simple technique to obtain very accurate eigenenergies for a confined hydrogen atom, as well as the critical-cage radii, to any desired degree of precision. The method is easily realized by the use of any contemporary mathematical software.  This makes the study of confined potentials more accessible.

\newpage
\section*{References}


\begin{thebibliography}{10}
\bibitem{Michels} A. Michels, J. de Boer and A. Bijl, Physica 4 (1937) 981.
\bibitem{colm} Jean-Patrick Correrade and Prasert Kengkan, {\it Atomic Confinement}, Ed. Colm T. Whelan and Nigel John Mason, Spring: New York 2005.
\bibitem{Sen} K. D. Sen, J. Chem. Phys. 122 (2005) 194324. 
\bibitem{Sinha} Anjana Sinha, Rajkumar Roychoudhury, and Y. P. Varshni, Can. J. Phys. 78 (2000) 141. 
\bibitem{Sommerfeld} A. Sommerfeld and H. Welker, Ann. Phys. N.Y. 32 (1938) 56.
\bibitem{froman} P.O. Fr\"oman, S. Yngve and N. Fr\"oman, J. Math. Phys. 28 (1987) 1813.
\bibitem{Jask} W. Jask\'olaski, Phys. Rep. 271 (1996) 1.
\bibitem{Buch} A. L. Buchachenko, J. Phys. Chem. B 105 (2001) 5839.
\bibitem{sch} A. V. Scherbinin, V. I. Pupyshev and A. Yu. Ermilov, in: Physics of Clusters, World Scientific, Singapore, 1997.
\bibitem{Pupyshev} V. I. Pupyshev and A. V. Scherbinin, Chem. Phys. Lett. 295 (1998) 271.
\bibitem{Scher}  A. V. Scherbinin and V. I. Pupyshev, Russ. J. Phys. Chem. 74 (2000) 292.
\bibitem{pupy} V. I. Pupyshev and A. V. Scherbinin, Phys. Lett. A 299 (2002) 371.
\bibitem{Zhu} J. L. Zhu, J. J. Xiong, and B. L. Gu, Phys. Rev. B 41 (1990) 6001.
\bibitem{Chuu} D. S. Chuu, C. M. Hsiao, and W. N. Mei, Phys. Rev. B 46 (1992) 3898.
\bibitem{Porras} N. Porras-Montenegro and S. T. Perez-Merchancano, Phys. Rev. B 46 (1992)
9780.
\bibitem{Parades} H. Parades-Guiterrez, J. C. Cuero-Yepez and N. Porras-Montenegro, J.
Appl. Phys. 75 (1994) 5150.
\bibitem{Varshni} Y. P. Varshni, Phys. Letters A 252 (1999) 248.
\bibitem{Gimarc}B. M. Gimarc, J. Chem. Phys. 44 (1966) 373.
\bibitem{Fernandez} F. M. Fernandez and E. A. Castro, Int. J. Quantum Chem. 21 (1982) 741.
\bibitem{Arteca} G. A. Arteca, F. M. Fernandez and E. A. Castro, J. Chem. Phys. 80 (1984)
1569.
\bibitem{david} D. Djajaputra and Bernard R. Cooper, Eur. J. Phys. 21 (2000) 281.
\bibitem{Krahmer} D. S. Kr\"ahmer, W P Schleich and V P Yakovlev, J. Phys. A: Math Gen. 31 (1998) 4493.
\bibitem{Goodfriend} P. L. Goodfriend, J. Phys. B: At. Mol. Opt. Phys. 23 (1990) 1373.
\bibitem{Marin} J. L. Marin and S. A. Cruz, Am. J. Phys. 56 (1988) 1134.
\bibitem{Cruz} J. L. Marin and S. A. Cruz, J. Phys. B: At. Mol. Opt. Phys. 24 (1991) 2899.
\bibitem{MarCruz} J. L. Marin and S. A. Cruz, Am. J. Phys. 59 (1991) 931.
\bibitem{ypv} Y. P. Varshni, J. Phys. B: At. Mol. Opt. Phys. 30 (1997) L589.
\bibitem{Guimaraes} M. N. Guimaraes and F. V. Prudente, J. Phys. B: At. Mol. Opt. Phys. 38 (2005) 2811.
\bibitem{Burrows} B. L. Burrows, M. Cohen, Int. J. Quant. Chem. 106 (2006) 478.
\bibitem{Var} Y. P. Varshni, J. Phys. B: At. Mol. Opt. Phys. 31 (1998) 2849.
\bibitem{Varsh} Y. P. Varshni, Z. Naturforsch. 57 (2002) 915.
\bibitem{Boeyens} J. C. A. Boeyens, J. Chem. Soc. Faraday Trans. 90 (1994) 3377.
\bibitem{Cecil} C. Laughlin, B. L. Burrows and M. Cohen, J. Phys. B: At. Mol. Opt. Phys. 35 (2002) 701. 
\bibitem{Ciftci} H. Ciftci, R. L. Hall, and N. Saad, J. Phys. A: Math. Gen. 36 (2003) 11807.
\bibitem{Hakan} H. Ciftci, R. L. Hall, and N. Saad, J. Phys. A: Math. Gen. 38 (2005) 1147. 
\bibitem{Abramowitz} M. Abramowitz and I. A. Stegun, {\it Handbook of Mathematical Functions}, Dover: New York, 1965.
\bibitem{champ} B. Champion, R. L. Hall and N. Saad, Int. J. Mod. Phys. A 23 (2008) 1405. 
\end{thebibliography}
\end{document}